\documentclass[12pt]{article}
\usepackage{graphicx}
\newcommand{\sss}{\scriptscriptstyle}

\newcommand {\be}{\begin{equation}} 
\newcommand{\ee}{\end{equation}}    

\def\dds1{\frac{\partial}{\partial s_1}}

\def\vtj{v_{{\sss T}j}}
\def\vta{v_{{\sss T}a}}
\def\vtb{v_{{\sss T}b}}

\def\d{d\kern-0.8 ex\vrule height 1.3 ex depth-1.24 ex width 0.7 ex
\kern 0.15 ex}
\def\D{D\kern-1.7 ex\vrule height .87 ex depth-0.8 ex width 0.7 ex
\kern 0.95 ex}

\begin{document}
\baselineskip 20 pt

\begin{center}

\Large{\bf Ion thermal  effects in oscillating multi-ion plasma sheath theory}

\end{center}

\vspace{0.5cm}

\begin{center}

J.~Vranjes$^{1,2}$, B.P.~Pandey$^{3}$, M.Y.~Tanaka$^{4}$, and S.~Poedts$^{1}$

\vspace{0.5cm}

{\em $^{1}$K.U. Leuven, Center for Plasma Astrophysics, and Leuven Mathematical Modeling and Computational Science Centre (LMCC),  Celestijnenlaan  200B, 3001 Leuven,  Belgium.
}

\vspace{5mm}

{\em $^{2}$Facult\'{e} des Sciences Appliqu\'{e}es, avenue F.D. Roosevelt 50,
1050 Bruxelles, Belgium.
}

\vspace{5mm}

  {\em  $^{3}$ Department of Physics, Macquarie University, Sydney, NSW
2109, Australia.}

\vspace{5mm}

{\em  $^{4}$Department of High Energy Engineering Science, Kyushu
University, Kasuga-koen 6-1, Kasuga, Fukuoka 816-8580, Japan
}

\end{center}

\vspace{2cm}

{\bf Abstract:} The effects of the ion temperature are discussed in a two-ion electron plasma and for a model applicable to the oscillating sheath theory that has recently been much in the focus of research. The differences between the fluid and kinetic models are pointed out, as well as the differences between the approximative kinetic description (which involves the expansion of the plasma dispersion function), and the exact kinetic description. It is shown that the approximative kinetic description, first, can not describe the additional acoustic mode which naturally exists in the plasma with an additional ion population with a finite temperature, and, second,  it yields an inaccurate  Landau damping of the bulk ion acoustic mode. The reasons for these two failures of the model are described. In addition to this, a fluid model is presented that is capable of capturing both of these features that are missing in the approximative kinetic description, i.e., two (fast and slow) ion acoustic modes, and the corresponding Landau damping of  both modes.

\vspace{0.8cm}


\pagebreak


\noindent{\bf I.~INTRODUCTION}

\vspace{0.5cm}

In the recent past there has been a lot of activity in the domain of theory and experiments dealing with the sheath in multi-ion plasmas.$^{1-5}$ An earlier work on this issue$^6$  provides us with a general expression for the Bohm criterion, while more recent theoretical$^3$  and experimental$^4$   works specify that, in a two-ion electron plasma, the sheath velocity of the two ion species have the same value that equals  the hybrid ion acoustic (IA) velocity. Yet, there seems to be no complete consensus about this in the literature.

The behavior of the ion acoustic mode in the presence of an additional ion species  may be drastically different compared to its behavior in a single ion plasma.$^{7-10}$ In particular, the Landau damping is increased with the addition of a lighter ion specie, which implies the possibility for controlling the mode behavior. Physically, this is equivalent to lowering the electron temperature and reducing the phase velocity, so that the damping is increased.$^7$ Experimental observations of those phenomena are available in the classic Refs.~8, 9,   and more recently in Ref.~10.   We observe  that these effects of the additional ions are just the opposite of the effects in the two (hot and cold) electron population case.$^{11,12}$

We stress that each additional ion specie introduces a new branch of acoustic oscillations in the system, complemented by various novel physical effects. However, in the context of the sheath theory, this domain of behavior of thermal ions has not yet been properly explored.
This is the subject of the present work. In the case of a plasma with  two ion species,  one obtains two acoustic modes: one fast and one slow mode. As a result,  the commonly used terms like the 'system sound velocity' appear to be redundant and may become inapplicable in most plasmas.

\vspace{0.5cm}

\noindent{\bf II. \,\,\, COMPARISON BETWEEN THE COLD  AND HOT IONS CASES}

\vspace{0.3cm}

\noindent{\bf A. Cold ions}
\vspace{0.3cm}

In order to see how thermal (kinetic) effects influence the oscillating plasma sheath, we first solve the dispersion equation for the
ion acoustic (IA)  mode propagating in a three component plasma consisting of electrons, and  two cold ion species denoted by $a$ and $b$:
\be
1+ \frac{1}{k^2} =\frac{1}{[\omega - k v_{a0} (n_{e0}/n_{a0})^{1/2} ]^2} + \frac{ (n_{b0}/n_{a0}) (m_a/m_b)}{[\omega  - k v_{b0}(n_{e0}/n_{a0})^{1/2}]^2}. \label{e1}
\ee
Here, similar to Ref.~1, we have $k\equiv k \lambda_{de}$, $\omega$ is normalized to $\omega_{pa}$, $v_{j0}\equiv v_{j0}/c_{sa}$ denotes the directed sheath currents/velocities of the two ion species, $j=a,b$, and  $c_{sa}^2= \kappa T_e/m_a$.

In the limit of large wave-lengths and for negligible sheath currents, with the given normalization, the dispersion properties of the hybrid bulk IA mode are described  by
\be
  \omega^2= k^2 [1+ m_a n_{b0}/(m_b n_{a0})].
   \label{aa}
   \ee
In physical units it reads $\omega^2/k^2= v_s^2\equiv (n_{a0}/n_{e0}) \kappa T_e/m_a + (n_{b0}/n_{e0}) \kappa T_e/m_b$, where $v_s$ denotes the hybrid IA sound speed in plasmas with two ion species, called 'the system sound speed' in the literature.$^3$

The equation~(\ref{e1}) for cold ions is solved numerically. First, similar to Ref.~1, we set $v_{a0}=0.009$, $v_{b0}=0.011$, which should correspond to the situation deep inside the plasma, far from the sheath, and  we also choose $m_a=10\, m_b$, and solve for $n_{b0}= 0.2 \,n_{a0}$   in terms of $k$. The result (for the ion acoustic mode) is presented in Fig.~1  with a dashed line, showing the tendency for the usual frequency saturation for large wave-numbers. Going to larger values of $k$ has no sense because of physical reasons. Two  additional complex-conjugate  solutions for the sheath current driven modes, that also follow from Eq.~(\ref{e1}),  are about 2 orders of magnitude lower and are not of interest here (see Section~III).

\vspace{0.5cm}

\noindent{\bf B. Hot ions}
\vspace{0.3cm}

Now, for comparison, keeping the ion thermal effects,  we re-derive the dispersion equation {\em using the kinetic theory}. It reads
\be
\triangle(\omega, k)\equiv 1+ \sum_\alpha (\omega_{p\alpha}^2/k^2 v_{{\sss T}\alpha}^2)\left[1- {\cal Z}(\omega_{\alpha 0}/kv_{{\sss T}\alpha})\right]=0.
\label{e2}
\ee
Here, $\alpha=e, a, b$, $\omega_{\alpha 0} =\omega - k v_{\alpha 0}$, and  ${\cal Z}(x)=[x/(2 \pi)^{1/2}] \int dy \exp(- y^2/2)/(x-y)$  is the plasma dispersion function, where  $x\equiv \omega_{\alpha 0}/(kv_{{\sss T}\alpha})$, $y \equiv v/v_{{\sss T} \alpha}$.
For non-streaming electrons and in the limit
\be
v_{{\sss T}a, b} \ll  \omega/k \ll  v_{{\sss T}e}, \label{con1}
\ee
the standard  expansions for   ${\cal Z}$ are used, and the general dispersion equation    (\ref{e2}) in that case becomes:
\[
1+ \frac{\omega_{pe}^2}{k^2 v_{{\sss T}e}^2} \left[1+ i (\pi/2)^{1/2} \frac{\omega}{k v_{{\sss T}e}} \right]
\]
\[
-
\frac{\omega_{p a}^2}{k^2 v_{{\sss T}a}^2} \left\{ \frac{k^2 v_{{\sss T}a}^2}{\omega_{a0}^2} +  \frac{3 k^4 v_{{\sss T}a}^4}{\omega_{a0}^4}
\right.
\left.
  -
i (\pi/2)^{1/2} \frac{\omega_{a0}}{k v_{{\sss T}a}} \exp\left[-\omega_{a0}^2/(2 k^2 v_{{\sss T}a}^2)\right]\right\}
\]
\be
-  \frac{\omega_{p b}^2}{k^2 v_{{\sss T}b}^2} \left\{ \frac{k^2 v_{{\sss T}b}^2}{\omega_{b0}^2} +  \frac{3 k^4 v_{{\sss T}b}^4}{\omega_{b0}^4}
\right.
\left.
   -
i (\pi/2)^{1/2} \frac{\omega_{b0}}{k v_{{\sss T}b}} \exp\left[-\omega_{b0}^2/(2 k^2 v_{{\sss T}b}^2)\right]\right\}=0.\label{e3}
\ee
In the appropriate limits its real part yields Eq.~(\ref{e1}), while the wave damping is given approximately by $\gamma \simeq - Im \triangle(\omega, k)/[\partial Re \triangle(\omega, k)/\partial \omega]_{\omega=\omega_r}$. Here, $Im \triangle(\omega, k)$ and $Re \triangle(\omega, k)$ denote the imaginary and real parts of Eq.~(\ref{e3}).

We have  solved Eq.~(\ref{e3}) numerically by  setting $T_a=T_b=T_e/15$,  and for the same $v_{j0}, n_{b0}$ as above. This result is also given in Fig.~1 (by full and dotted lines).  The Landau damping has the maximum at $k\simeq 1$ where $|\gamma|/\omega\simeq 0.25$. Note that here and further in the text we present the {\em absolute} value of the Landau damping.

Observe a considerable change in the mode frequency. This  may be of even  greater importance than the damping because, compared to the cold ion limit,   the phase speed is increased, which should be taken into account in modeling  the Bohm criterion.

The presence of the second ion species affects both the mode frequency and the Landau damping, as  shown in Fig.~2, where the hybrid IA mode frequency (normalized to $\omega_{pa}$)   is presented in terms of the  number density of the species $b$, and for $T_a=T_b=T_e/30$, and $k=0.1$ (cf.\ Fig~1). Here, in the beginning the Landau damping is increased by increasing the amount of the ion species $b$, but this goes only up to $n_b/n_a$ of about 6 percent when $|\gamma|/\omega\simeq 0.11$. Note that for $n_{b0}=0$ we have  $|\gamma|=2\cdot 10^{-4}$ (in units of $\omega_{pa}$), i.e., an 80 times lower damping compared with  the case when $n_{b0}/n_{a0}=0.06$!  Clearly, the presence of the additional  ion species significantly affects the damping of the mode. This is because the wave-particle exchange of energy is easily facilitated when additional ions are available in the system.

For hotter ions (i.e., $T_a=T_b=T_e/10$), the ratio $|\gamma|/\omega$ is increased, with the maximum $\simeq 0.22$ at $n_{b0}/n_{a0}=0.2$, as seen from Fig.~3. Here, the Landau damping grows with the addition of more $b$-ions, yet this increase in the damping  saturates  for  the number density $n_b$ reaching about 20 percent.  Note that for $n_{b0}=0$  it is  $|\gamma|=4\cdot 10^{-3}$, which is 12.5 lower compared with the case at  $n_{b0}/n_{a0}=0.2$.

Hence, the second ion species may drastically increase the Landau damping of the hybrid IA mode, and we see also  that its {\em relative}  effect is more pronounced  for lower ion temperatures.

\vspace{0.5cm}

\noindent{\bf III. SHEATH-CURRENT DRIVEN MODES}

\vspace{0.5cm}

As is well known,$^{13}$ the presence of directed fluxes of plasma components   implies  the presence of extra collective modes. In our case  these are  the ion sheath-current (SC) driven,  complex-conjugate solutions with frequencies of the order of $\approx k v_{j0}$,  and with a growth rate that is approximately one order of magnitude lower. Their presence is obvious from the form of Eq.~(\ref{e1}) and they are discussed below.

Increasing the directed ion velocities, the IA mode behavior is changed. This is visible from Fig.~4,  where  the velocities $v_{j0}$ are multiplied  by a factor 100, becoming  close to the values expected for the Bohm criterion,  and for  $n_b=0.2 n_a$, $T_a=T_b=T_e/10$.
Here, the normally negative solution for the hybrid  IA mode (that is not discussed above in Fig.~1 as being of no interest)  is so strongly Doppler shifted that it becomes positive (line $b$). The absolute value of the  Landau damping (the full line in Fig.~5) reaches its maximum  at $k\simeq 0.7$ when $|\gamma|/\omega_r=0.16$.

On the other hand, the growing part of the two complex-conjugate SC solutions, in the $k$ domain presented in Fig.~5, has  a growth rate that  is about one order of magnitude below the mode frequency. It can easily be shown that for even  larger values of $k\geq 15$ the instability vanishes and the real part of the SC mode frequency splits into two modes. However, this implies wavelengths below the electron Debye length and the model has no sense in that limit.

\vspace{0.5cm}

\noindent{\bf IV. \,\,\, ADDITIONAL ACOUSTIC MODE}

\vspace{0.5cm}

Apart from the demonstrated Landau damping and the modification of the frequency, the additional hot ion species further imply additional branches of acoustic oscillations. This can only be seen by solving the general dispersion equation (\ref{e2}) numerically, like in Refs.~7, 9.
Formally, the additional acoustic branches appear in the limit when  the terms $k^2 \vtj^2$ are kept finite in comparison with $\omega^2$. Therefore, in view of the condition (\ref{con1}), these additional acoustic branches are absent in the previous analysis and can not be deduced from Eq.~(\ref{e3}). Hence, the approximative kinetic approach  describes the Landau damping but it is unable to describe the {\em  additional} acoustic branch. However, a recently developed fluid theory$^{14}$ easily captures both of these features within the fluid theory. For Boltzmannian electrons and with the help of the Poisson equation, the derivation of the dispersion equation   is straightforward yielding:$^{14}$
\be
1+ \frac{1}{k^2 \lambda_{de}^2} = \frac{\omega_{pa}^2}{\omega_{a0} \omega_{a1}- k^2 \vta^2} + \frac{\omega_{pb}^2}{\omega_{b0} \omega_{b1}- k^2 \vtb^2}.  \label{f}
\ee
Here, $\omega_{j1}= \omega_{j0} + i \mu_{j0} k^2$. In the cold ions limit and for the given normalization, Eq.~(\ref{f}) becomes identical to Eq.~(\ref{e1}). The term $\mu_{j0}\equiv\lambda v_{sj}/(2 \pi^2 d_j)$, where $v_{sj}^2= c_{sj}^2 + \vtj^2$,  follows from the ion momentum equation of the form
\[
(\partial/\partial t + \vec v_{j0} \nabla) v_{j1} = - (q_j/m_j)\partial \phi_1/\partial x - (\kappa T_j/n_{j0}) \partial n_{j1}/\partial x +
\mu_{j0} \partial^2 v_{j1}/\partial x^2,
  \]
where it has been introduced to describe the Landau damping. Such a fluid model has first been used in Ref.~15, and more recently in Refs.~14,~16.
It is convenient because it allows the use of the fluid theory, where the expressions can be analyzed more easily, as compared to the general kinetic expression (\ref{e2})  containing the plasma dispersion function expressed through the integral ${\cal Z}(x)=[x/(2 \pi)^{1/2}] \int dy \exp(- y^2/2)/(x-y)$. The term $d_j\equiv\delta_j/\lambda$ in $\mu_{j0}$ gives the ratio of the Landau attenuation length $\delta_j$ and wave-length $\lambda$. It is chosen in such a way that it is independent of the wavelength and the plasma density, and it depends on the ion temperature in a prescribed way. As is shown in Refs.~14,~16, such a fluid model for the intrinsically  kinetic Landau damping is, first,  {\em much more accurate} than the approximative kinetic Landau damping obtained after the expansion of ${\cal Z}(x)$ over the parameter $x$.
Second, it gives {\em the same damping} as the exact kinetic Landau term.   Third, it yields the analytical expression for the acoustic modes in two-ion electron plasmas with different temperatures of the two ion species [the alternative is numerically solving Eq.~(\ref{e2})].

All these features  follow after adopting  the following expression:$^{14,16}$ 
\be
d_j\equiv \delta_j/\lambda \approx 0.2750708 + 0.0420737789\, \tau_j +
    0.0890326 \,\tau_j^2 - 0.011785 \, \tau_j^3 +
    0.0012186\,\tau_j^4.  \label{ff}
\ee
Simple derivations$^{14}$  for an electron ion plasma yield the modeled fluid Landau damping as:
\be
|\gamma_f|/\omega= 1/(2 \pi d). \label{fl}
\ee
The approximative$^{17}$ kinetic Landau damping is:
\be
|\gamma_{app}|/\omega= (\pi/8)^{1/2}\left[(m_e/m_i)^{1/2} + \tau (3+ \tau)^{1/2} \exp \left[-(3+ \tau)/2\right]\right].
\label{app}
\ee
Finally, the exact kinetic Landau damping  can be obtained only numerically. However, using the graph of the exact kinetic Landau damping from Ref.~11 one finds$^{14}$ that it  may be  expressed by the following polynomial:
  \[
|\gamma_{ex}|/\omega=0.681874545  -0.369763643   \tau +
  0.0934588855 \tau^2
  \]
  \be
  - 0.01203427 \tau^3+
  0.0007523967677 \tau^4 -0.000018 \tau^5.\label{ex}
  \ee
The above given statements about the accuracy of the expressions (\ref{fl},\ref{app}) can directly be checked by plotting the expressions (\ref{fl}-\ref{ex}) in terms of $\tau$, see Fig.~2 from Ref.~14.

It appears that, in the case of a two-ion electron plasma, the behavior of two acoustic modes is determined by the relation between $c_{sa}$ and $\vtb$. More details are available in Refs.~9, 14.  Eq.~(\ref{f})  gives two (fast and slow) ion acoustic modes (that appear due to the fact that ions have some temperature),  together with the corresponding  Landau damping. As will be demonstrated below, this raises the question about the appropriateness of the 'system sound velocity', that is commonly used in the literature.

Using the same normalization as earlier, Eq.~(\ref{f}) becomes
\[
1+ \frac{1}{k^2}=\left\{\left(\omega- k v_{a0} \sqrt{n_{e0}/n_{a0}}\right) \left[\omega- k v_{a0} \sqrt{n_{e0}/n_{a0}} + i [k v_{sa}/(\pi d_a)]\sqrt{n_{e0}/n_{a0}}\right]
\right.
\]
\[
\left.
- k^2 T_a n_{e0}/(T_e n_{a0})\right\}^{-1}
\]
\[
+
\frac{n_{b0}}{n_{a0}} \frac{m_a}{m_b} \left\{ \left(\omega- k v_{b0} \sqrt{n_{e0}/n_{a0}}\right)
\left[\omega- k v_{b0} \sqrt{n_{e0}/n_{a0}} + i [k  v_{sb}/(\pi d_b)]\sqrt{n_{e0}/n_{a0}}\right] \right.
\]
\be
\left.
- k^2 m_a n_{e0} T_b/(m_b n_{a0} T_e)\right\}^{-1}. \label{f2}
\ee
Here, $v_{sj}^2\equiv (c_{sj}^2 + \vtj^2)/c_{sa}^2$. For comparison with previous cases, we solve Eq.~(\ref{f2}) by taking the same parameters as in Fig.~1, i.e., $n_{b0}=0.2 n_{a0}$, $v_{a0}=0.009$, $v_{b0}=0.011$, $v_{sa}=1.03$, $v_{sb}=3.26$, $m_a=10 m_b$. We have also chosen $T_e=11600$ K, $T_a=T_b=T_e/15$, and $m_b=4 m_p$ so that $c_{sa}=1547\;$m/s. The two acoustic modes are presented in Fig.~6.  The given sheath currents make  negligible frequency shifts in the two acoustic modes. The corresponding Landau damping of the two modes, presented in Fig.~7, shows an obvious difference  as compared to Fig.~1. It is, first, lower by magnitude and, second, it shows no decrease for larger values of $k$. Hence, the decrease and saturation of the Landau damping,  seen in  previous figures, is clearly only due to the expansion of the plasma dispersion function, i.e., an artefact of  the approximation used,  and may not have any relevance to the real physical systems.

Similarly, when $n_{b0}$ is increased the frequency and the Landau damping of the  upper (fast) IA mode is increased. However, for the slow IA mode  these both parameters are  reduced when $n_{b0}$ is increased. This is checked by setting $k=0.3$ and for other parameters as above. The results are  presented in Fig.~8 (for the two acoustic frequencies), and Fig.~9 for the absolute value of the Landau damping (multiplied by $10^3$). This behavior is in agreement with  results existing in the literature.$^{9,14}$

Setting the ion temperature to a larger value,  i.e., $T_a=T_b=T_e/2$, the frequencies are increased by about a factor 1.5. However, the Landau damping is increased by a factor 100. This is presented in Figs.~10 and 11. It is seen also that the damping of the slow mode is now increased for larger values of $n_{b0}$.

\vspace{0.5cm}

\noindent{\bf V. \,\,\, SUMMARY}

\vspace{0.5cm}

To summarize, the ion thermal effects are investigated in the sheath theory in multi-component plasmas containing
two ion species and electrons. Typically, this implies using the kinetic theory and we have demonstrated the differences between some results existing in the literature (obtained  without ion thermal effects), and those obtained in the present analysis when the ion temperature effects are taken into account. The most important additional effect which follows from the finite ion temperature  is the Landau damping, that is discussed in detail. However, the standard analysis implies the expansion of the plasma dispersion function, and as a result a)~some phenomena related to ion thermal effects are missed in the procedure, and b)~the mode behavior in some limits may show features that are not physical and that are only due to the mentioned expansion. The most obvious example of a) is the presence of an additional acoustic mode that can be seen only by solving the general kinetic dispersion equation (\ref{e2}) numerically.$^{7,9}$ As for b), such an artefact of the expansion is the saturation and decrease of the Landau damping, as demonstrated in Secs. 2,~3.
In the case of hot ions, the expansion of the plasma dispersion function is not justified and an alternative approach is needed. We have demonstrated in Sec.~4 that such a method exists within the framework of fluid theory.$^{14-16}$ It is simple and suitable for analytical work,  and at the same time is  capable of capturing both effects, the additional ion acoustic mode and the Landau damping. In our earlier works$^{14,16}$ we have shown that, in fact, quantitatively it describes the Landau damping much more accurately than the analysis which follows after the expansion of the plasma dispersion function.

We conclude that more care is needed whenever the standard expansion of the plasma dispersion function is used. This may  directly be seen by comparing figures from Sections~2 and 3 on one side, and those from Sec.~4 on the other side. We stress also that even relatively  small ion thermal effects imply the presence of an extra (slow) acoustic branch of ion oscillations, so that in fact two acoustic modes propagate in the plasma. The examples given in Sec.~4 show that the frequencies of the two modes are not very distant even for relatively cool ions (in the given case the ion temperature of only about $1/15$ of the electron temperature), yet the two modes have rather different behavior. Therefore, terms like 'system sound velocity' may become inapplicable for most plasmas.

A possible extension of the results presented here could be the  case of a two-ion electron plasma with one {\em negative} ion specie, and  in particular the pair-ion electron plasma produced recently in the laboratory$^{18}$ and discussed in the literature very extensively.$^{19-23}$

\vspace{1cm}

\paragraph{Acknowledgements:}
The  results presented here  are  obtained in the framework of the
projects G.0304.07 (FWO-Vlaanderen), C~90205 (Prodex),  GOA/2004/01
(K.U.Leuven),  and the Interuniversity Attraction Poles Programme -
Belgian State - Belgian Science Policy. The work has been completed  during the visit of JV to the Department of High Energy Engineering Science, Kyushu University. The hospitality and excellent working conditions are gratefully acknowledged.

\pagebreak

\pagebreak

\noindent {\bf Figure captions:}
\begin{description}
\item{Fig. 1:} The positive solutions for the hybrid IA mode frequency (normalized to $\omega_{pa}$) for hot (full line), and cold ions (dashed line), and the absolute value of the normalized Landau damping $|\gamma|$ (dotted line).

\item{Fig. 2:} Full and dashed lines: the positive frequency (normalized to $\omega_{pa}$) of the hybrid IA mode  in terms of the number density of  the second ion species, for $k=0.1$. Dotted line:  the absolute value of the normalized Landau damping $|\gamma|$.

\item{Fig. 3:} The same as in Fig.~2 but for hotter ions $T_a=T_b=T_e/10$.

\item{Fig. 4:} The positive IA  mode (line a), and its normally negative counterpart,  here strongly  positively Doppler-shifted   (line b). Dashed line: the sheath-current (SC)
mode. Here $k\equiv k \lambda_{de}$.  

\item{Fig. 5:} Full line: the absolute value of the IA Landau damping $|\gamma|$ (corresponding to Fig~4); dashed line:  the corresponding growth rate of the SC mode.

\item{Fig. 6:} Two ion acoustic modes in two-ion electron plasma with $T_a=T_b=T_e/15$ (other parameters given in the text). 

\item{Fig. 7:} The absolute Landau damping $|\gamma|\equiv |\gamma|/\omega_{pa}$  (multiplied by $10^{2}$) of the two IA modes from Fig.~6.   

\item{Fig. 8:} The frequencies of the two ion acoustic modes in terms of the number density of the ion species $b$ for  $T_a=T_b=T_e/15$.     

\item{Fig. 9:} The absolute Landau damping $|\gamma|\equiv |\gamma|/\omega_{pa}$  (multiplied by $10^{3}$) of the two IA modes from Fig.~8.  

\item{Fig. 10:} The frequencies of the two ion acoustic modes in terms of the number density of the ion species $b$ for $T_a=T_b=T_e/2$.  

\item{Fig. 11:} The absolute Landau damping $|\gamma|\equiv |\gamma|/\omega_{pa}$  of the two IA modes from Fig.~10 (for $T_a=T_b=T_e/2$).    

\end{description}


\begin{thebibliography}{99}
\bibitem{f} R. N. Franklin, J. Phys. D: Appl.  Phys. {\bf 36}, 1806 (2003).
\bibitem{w} X. Wang and   N. Hershkowitz, Phys. Plasmas  {\bf 13}, 053503 (2006).
\bibitem{lee} D. Lee, L. Oksuz, and N. Hershkowitz,  Phys. Rev. Lett. {\bf 99}, 155004 (2007).
\bibitem{lee2} D. Lee, N. Hershkowitz, and G. D. Severn, Appl. Phys.  Lett. {\bf 91}, 0411505 (2007).
\bibitem{ok} L. Oksuz, D. Lee, and  N. Hershkowitz, Plas. Sources Sci. Tech. {\bf 17}, 015012 (2008).
\bibitem{rie} K. U. Riemann, IEEE Trans. Plas. Sci. {\bf 23}, 709 (1995).
\bibitem{fr} B. D. Fried, R. B. White, and T. K. Samec, Phys. Fluids  {\bf 14}, 2388 (1971).
\bibitem{al} I. Alexeff, W. D. Jones, and D. Montgomery,  Phys. Rev. Lett. {\bf 19}, 422 (1967).
\bibitem{nak} Y. Nakamura, M. Nakamura, and T. Itoh, Phys. Rev. Lett. {\bf 37}, 209 (1976).
\bibitem{nak2} Y. Nakamura and Y. Saitou, Plasma Phys. Control. Fusion {\bf 45}, 759 (2003).
\bibitem{ch} F. F. Chen, {\em Introduction to Plasma Physics and Controlled Fusion} (Plenum, New York,  1984) p. 272.
\bibitem{vr2} J. Vranjes and S. Poedts, Eur. Phys. J. D. {\bf 40}, 257  (2006).
\bibitem{is} S. Ichimaru,    {\em Basic Principles of Plasma Physics} (The Benjamin/Cummings Publish. Comp., Reading, Massachusetts, 1973) p. 144.
\bibitem{vr1} J. Vranjes, M. Y. Tanaka, and S. Poedts, Phys. Plasmas {\bf 13}, 122103 (2006).
\bibitem{da} N. D'Angelo, G. Joyce, and M. E. Pesses, Astrophys. J. {\bf 229}, 1138 (1979).
\bibitem{vr3} J. Vranjes, B. P. Pandey, and S. Poedts, Phys. Plasmas {\bf 14}, 032106 (2007).
\bibitem{gol} R. J. Goldston and P. H. Rutherford,  {\em Introduction to Plasma Physics} (Institute of Physics, Bristol,   1995) pp. 446-448.
\bibitem{h1} W. Oohara and R. Hatakeyama, Phys. Rev. Lett. {\bf
91}, 205005 (2003).
\bibitem{v4} J. Vranjes and S. Poedts, Plas. Sources Sci. Tech. {\bf 14}, 485  (2005).
\bibitem{s6} A. Luque, H. Schamel, B. Eliasson, and P. K. Shukla, Phys. Plasmas {\bf  12} 122307 (2005).
\bibitem{k1}  I. Kourakis, A. Esfandyari-Kalejahi, M. Medhipoor, and P. K. Shukla,  Phys. Plasmas  {\bf 13}, 052117 (2006).
\bibitem{s2} H. Saleem, Phys. Plasmas  {\bf 14}, 014505 (2007).
\bibitem{v5} J. Vranjes, D. Petrovic, B. P. Pandey, and S. Poedts, Phys. Plasmas {\bf 15}, 072104  (2008).
\end{thebibliography}
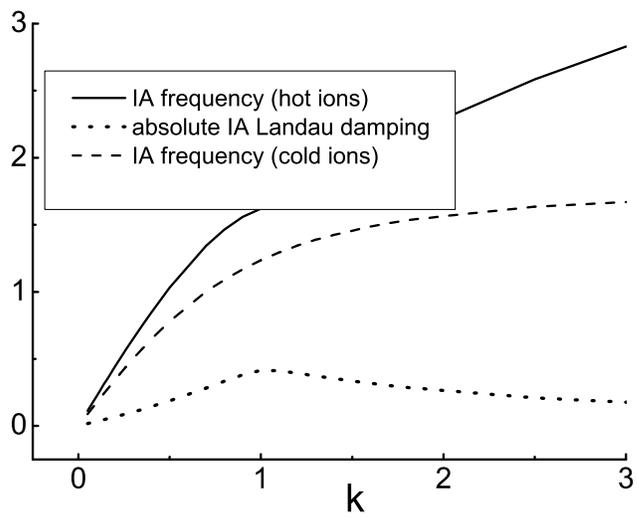
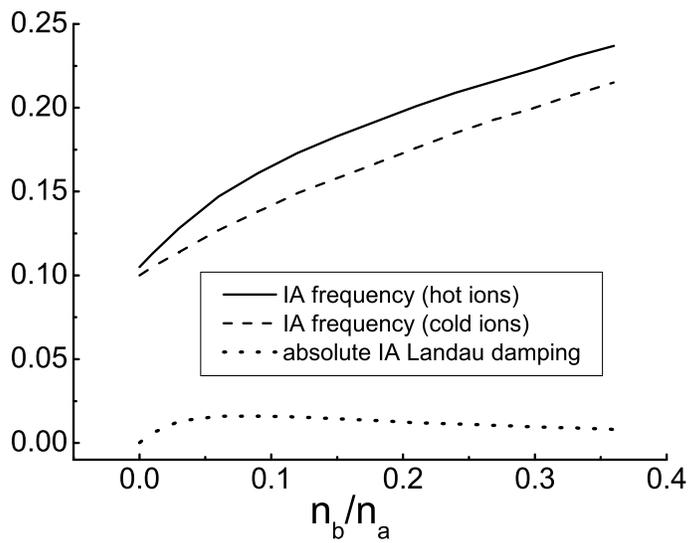
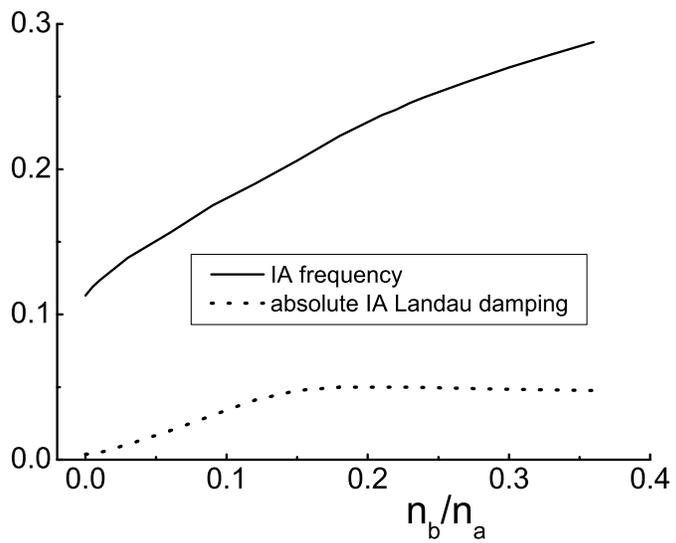
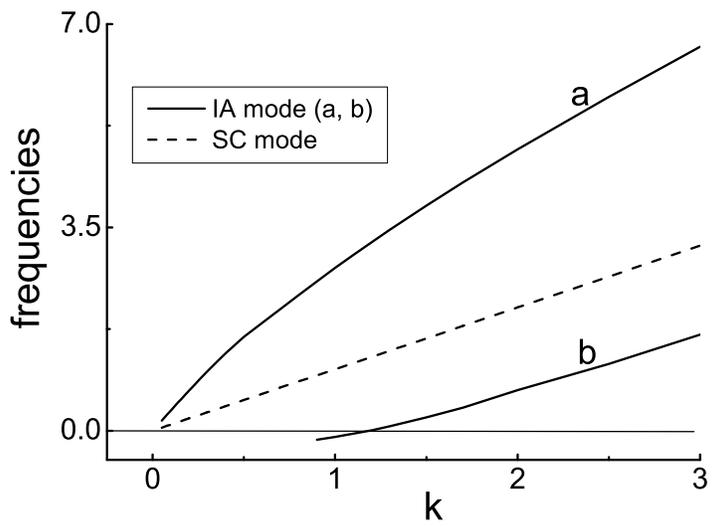
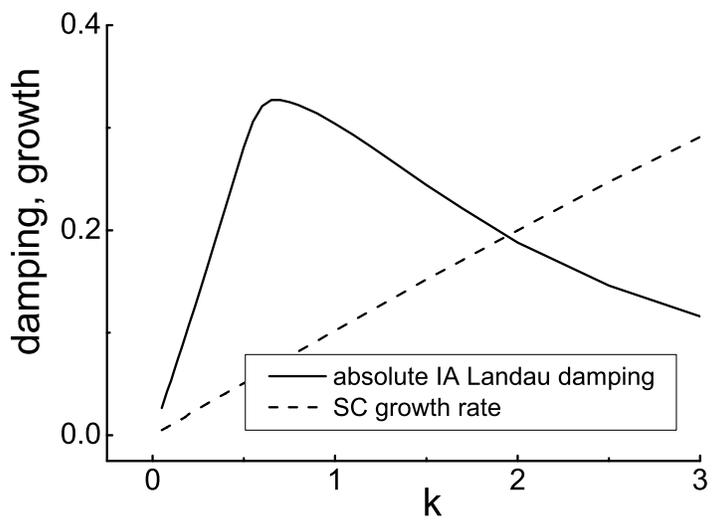
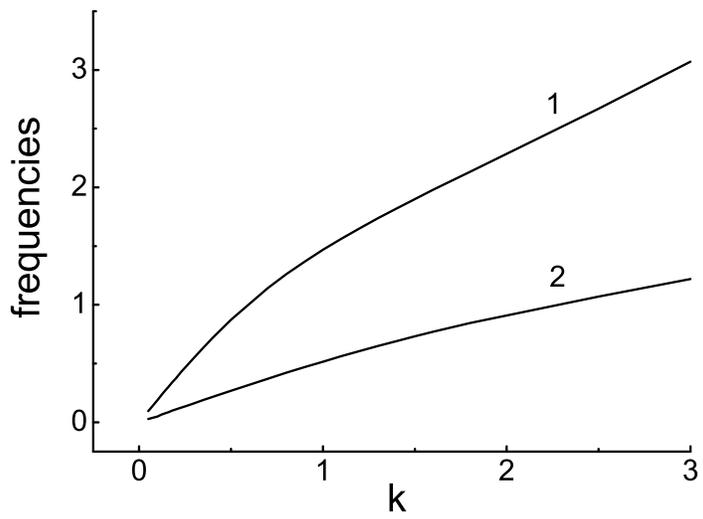
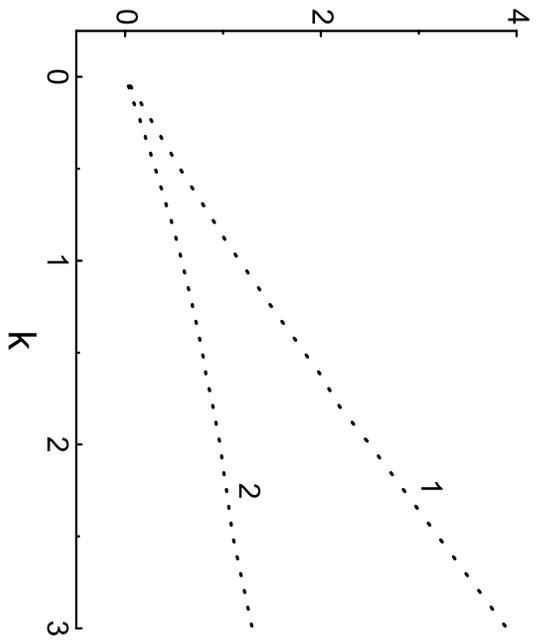
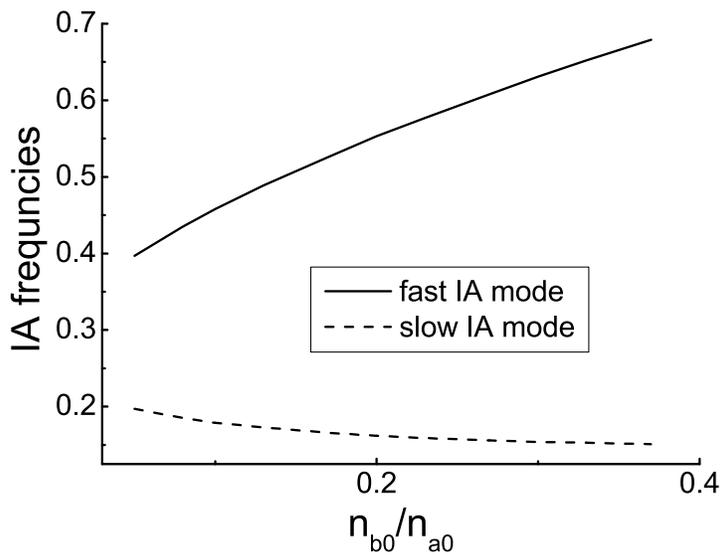
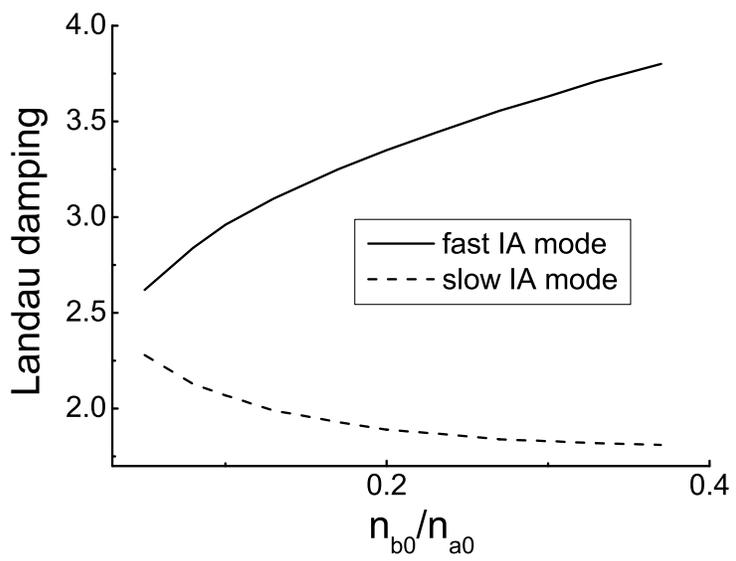
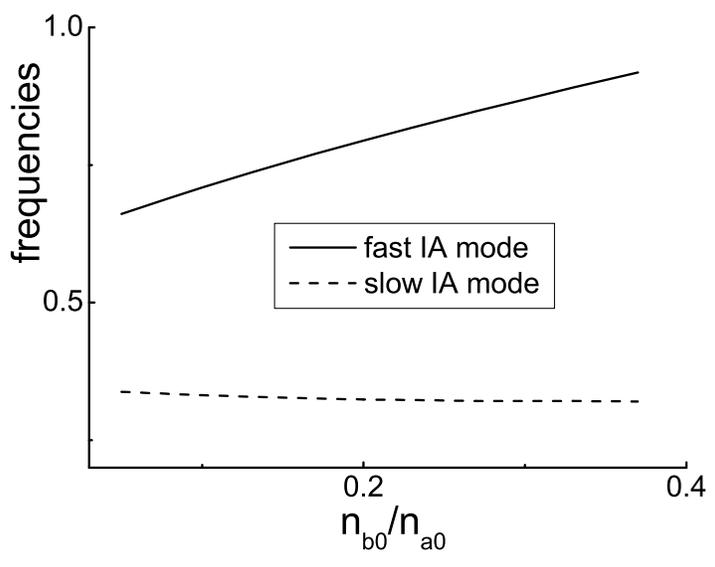
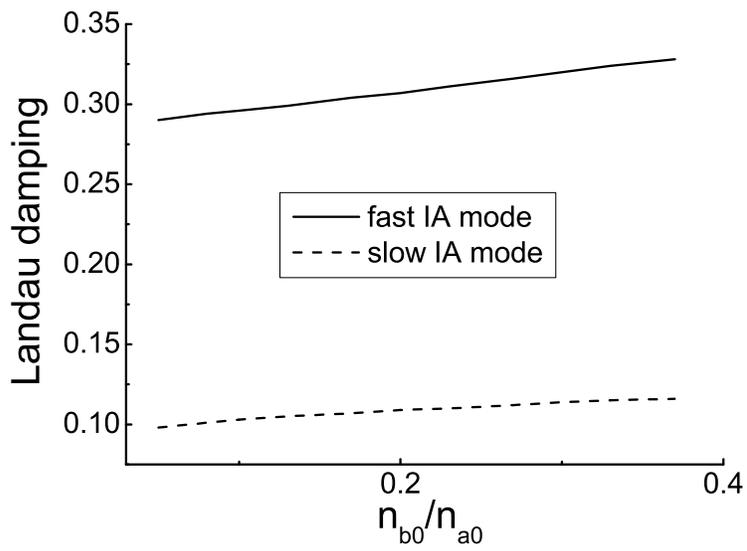
\end{document}